\documentclass[pra,amsmath,aps,twocolumn,superscriptaddress,letterpaper]{revtex4}
\usepackage{graphicx,xspace,subfigure}
\usepackage{float}
\usepackage{amsmath}
\usepackage{amssymb}
\usepackage[normalem]{ulem}
\usepackage{wrapfig}
\usepackage{epsfig}
\usepackage{color}
\input{epsf}

\definecolor{BrickRed}{cmyk}{0,0.89,0.94,0.28}
\definecolor{MidnightBlue}{cmyk}{0.98,0.13,0,0.43}
\definecolor{DarkGreen}{rgb}{0,0.7,0.1}

\newcommand{\vecs}{\mathbf{s}}
\newcommand{\vecsa}{\mathbf{s}_\alpha}
\newcommand{\vecsb}{\mathbf{s}_\beta}
\newcommand{\vecsperp}{\vecs^\bot}
\newcommand{\vecu}{\mathbf{u}}
\newcommand{\vecx}{\mathbf{x}}

\newcommand{\veck}{\mathbf{k}}

\newcommand{\tM}{\tilde{M}}
\newcommand{\mtM}{$\tM$ }
\newcommand{\tMinfi}{\tM_\infty^{-1}}

\newcommand{\vecseparation}{\mathbf{a}}
\newcommand{\Imatrix}{\mathbf{1}}
\newcommand{\mbra}{\langle m |}

\newcommand{\mpket}{|m'\rangle}

\newcommand{\qtwobra}{\langle q_2|}

\newcommand{\qtwopket}{|q_2'\rangle}
\newcommand{\inv}{^{-1}}
\newcommand{\sX}{s^X}
\newcommand{\ZX}{Z^X_{m,m'}}
\newcommand{\onep}{_\text{1p}}
\newcommand{\tMonep}{\tM\onep}
\newcommand{\tMonepinfi}{\tM_{\text{1p},\infty}^{-1}}
\newcommand{\twop}{_\text{2p}}
\newcommand{\tMtwop}{\tM\twop}
\newcommand{\np}{_\text{np}}
\newcommand{\tMnp}{\tM\np}

\newcommand{\EPFA}{E_\text{PFA}}
\newcommand{\EPFAcylplate}{\EPFA^\text{cyl-plate}}
\newcommand{\EPFAcylcyl}{\EPFA^\text{cyl-cyl}}

\begin{document}

\title{Casimir forces between cylinders and plates}
\author{Sahand Jamal Rahi}
\affiliation{Massachusetts Institute of Technology, Department of
  Physics, 77 Massachusetts Avenue, Cambridge, MA 02139, USA}
\author{Thorsten Emig}
\affiliation{Institut f\"ur Theoretische Physik, Universit\"at zu
  K\"oln, Z\"ulpicher Strasse 77, 50937 K\"oln, Germany}
\affiliation{Laboratoire de Physique Th\'eorique et Mod\`eles
  Statistiques, CNRS UMR 8626, B\^at.~100, Universit\'e Paris-Sud, 91405
  Orsay cedex, France}
\author{Robert L. Jaffe}
\affiliation{Massachusetts Institute of Technology, Department of
  Physics, 77 Massachusetts Avenue, Cambridge, MA 02139, USA}
\affiliation{Center for Theoretical Physics, Laboratory for Nuclear Science,
Massachusetts Institute of Technology, Cambridge, MA 02139, USA}
\author{Mehran Kardar}
\email{kardar@mit.edu}
\affiliation{Massachusetts Institute of Technology, Department of
  Physics, 77 Massachusetts Avenue, Cambridge, MA 02139, USA}

\begin{abstract}   
  We study collective interaction effects that result from the
  change of free quantum electrodynamic field fluctuations by
  one- and two-dimensional perfect metal structures. The Casimir
  interactions in geometries containing plates and cylinders is
  explicitly computed using partial wave expansions of constrained path
  integrals. We generalize previously obtained results and provide a
  more detailed description of the technical aspects of the approach
  \cite{Emig06}. We find that the interactions involving cylinders
  have a weak logarithmic dependence on the cylinder radius,
  reflecting that one-dimensional perturbations are marginally
  relevant in 4D space-time. For geometries containing two cylinders
  and one or two plates, we confirm a previously found non-monotonic
  dependence of the interaction on the object's separations which does
  not follow from pair-wise summation of two-body
  forces. Qualitatively, this effect is explained in terms of
  fluctuating charges and currents and their mirror images.
\end{abstract}
\maketitle

\section{Introduction}

Quantum effects like Casimir forces have become increasingly important
as electronic and mechanical systems on the nanometer scale become
more prevalent \cite{Cleland96, Chan01}. Now mechanical oscillation
modes of quasi one-dimensional structures such as nano wires or carbon
nanotubes can be probed with high precision \cite{Sazonova04}.
Generally, the behavior of such systems is influenced by the
collective nature of fluctuation forces: The total interaction of a
system of objects or particles cannot be obtained by simply adding the
forces between all pairs. Instead one has to consider also 3-body
and higher order interactions that become increasingly important
with decreasing separations between the objects. 

So far Casimir electrodynamic interactions have mostly been
investigated for two objects: parallel plates \cite{Casimir48-1}, a
rectilinear piston \cite{Hertzberg05}, plate-sphere interaction at
asymptotically large distances \cite{Casimir48-2} and for all
separations only recently \cite{Emig:2007a}. A previous letter
summarized the results for a plate and a parallel cylinder
\cite{Emig06}.  While shape and geometry can strongly influence
two-body Casimir interactions, it is also important to understand the
consequences of the non-additivity of fluctuation forces. In addition,
the extent to which fluctuations are correlated depends on the
effective dimensionality of the space that can be explored by the
fluctuations. Therefore, Casimir interactions are expected to depend
strongly on the codimension of the interacting objects
\cite{Scardicchio:2005b}.

In this work we concentrate on two central aspects of fluctuation
forces: Effects resulting from the non-additivity and the particular
properties of systems with a codimension of the critical value of
two. We consider these problems in the context of interactions between
cylinders and sidewalls. In previous works we have demonstrated that
Casimir forces in these geometries have only a weak logarithmic
dependence on the cylinder radius \cite{Emig06} and can be
non-monotonic \cite{Rodriguez07:PRL,Rahi:2008mz} -- consequences of
codimension and non-addivity.  Here we employ and extend previously
developed methods \cite{Emig06} to obtain the exact interaction
between two perfect metal cylinders and its modification due to
sidewalls from a partial wave expansion.  These geometries are of
recent experimental interest since cylinders are easier to hold
parallel and generate a force that is extensive in its length
\cite{Brown-Hayes05}.

In analogy to the cylinder-plate interaction \cite{Emig06}, we obtain
a weak logarithmic dependence on the cylinder radii $R_\alpha$ for the
interaction $E \sim - \hbar c L/[d^2 \log(d/R_1)\log(d/R_2)]$ between
two parallel cylinders at asymptotically large distance $d\gg R$.  We
include higher order partial waves to describe the crossover between
the asymptotic expression and the interaction at very short
separations where the proximity force approximation (PFA) gives the
correct zeroth order approximation to the Casimir energy. For two
cylinders of equal radius $R$, it has the form
$\EPFAcylcyl=-\frac{\pi^3}{1920} \hbar c L\sqrt{R/(d-2R)^5}$
\cite{Rahi:2008mz}.  When one or two perfectly conducting sidewalls
are added to a pair of cylinders, the force between the cylinders
depends non-monotonically on the sidewall separation $H$
\cite{Rahi:2008mz}. We compute the interaction between the cylinders
and the cylinder and the sidewall over a wide range of separations by
employing the method of images and by summing numerically a large
number of partial wave contributions. The non-monotonic behavior is
found to result from a competition between force contributions from
transverse magnetic (TM) and electric (TE) modes which induce opposite
image sources.
The TE and TM forces between two cylinders are monotonically
increasing and decreasing with the separation of the plate,
respectively; their sum behaves non-monotonic because the slopes are
different.

The rest of the paper is organized as follows. In the following
Section we describe the methodology of our path integral approach and
derive the elements of the relevant matrix operators for cylinders and
sidewalls in a partial wave basis. In Section \ref{sec:int_energies}
we obtain analytical and numerical results for the forces
between cylinders and sidewalls. Experimental implications
and corrections for cylinders of finite length are discussed
in Section \ref{sec:discussion}. More technical steps of the
calculations are relegated to the Appendices.

\section{Methods}
\subsection{Path integral and partial waves}

We consider geometries that are composed of infinitely long cylinders
and infinitely extended plates that are oriented such that the
cylinder axes are all parallel and coincide with an in-plane axis of
the plates that are also parallel to each other. Hence, the geometries
have one continuous translational symmetry. This allows the
electromagnetic modes to be split into transversal magnetic (TM)
modes, described by a scalar field with Dirichlet (D) boundary
conditions, and transversal electric (TE) modes, described by a scalar
field obeying Neumann (N) boundary conditions. After a Wick rotation
to the imaginary frequency ($q_0$) axis, the action for the scalar
field has the simple form
\begin{equation}
\label{eq:action}
S=\frac{1}{2}\int dq_0 \int d^3x
  (|\nabla\Phi|^2 + q_{0}^{2}\Phi^{2}) \, . 
\end{equation}
For the implementation of the boundary conditions and the computation
of the interaction energy, we employ the techniques derived in
references \cite{Li91,Li92,Buescher05,Emig06}. After introducing an
auxiliary field on each boundary to enforce the boundary conditions
and integrating out the scalar field $\Phi$, one obtains an effective
quadratic action for the auxiliary fields with kernel
\begin{equation}
  \label{eq:D-kernel}
  M_{\alpha \beta}(\vecu,\vecu';q_0) = G_0(\vecsa(\vecu),\vecsb(\vecu');q_0)
\end{equation}
for D conditions and
\begin{equation}
  \label{eq:N-kernel}
  M_{\alpha \beta}(\vecu,\vecu';q_0) = \partial_{\mathbf{n}_\alpha (\vecu)} \partial_{\mathbf{n}_\beta
(\vecu')} G_0(\vecsa(\vecu),\vecsb(\vecu');q_0)
\end{equation}
for N conditions, where the indices $\alpha$, $\beta$ label the
surfaces. Here $G_0(\vecx,\vecx';q_0)=e^{-q_0 |\vecx-\vecx'|}/4 \pi
|\vecx-\vecx'|$ is the free space Green's Function and $\vecsa(\vecu)$
is a vector pointing to the $\alpha$'th surface parametrized by the
coordinate vector $\vecu$ which describes a surface in 3D, so that it
stands for two independent parameters, e.g., for a cylinder
$\vecu=(x_1,\phi)$ with $x_1$ oriented parallel the cylinder axis and
$\phi$ the azimuthal angle.  When we integrate over the auxiliary
fields, we finally obtain the Casimir energy for D and N modes at zero
temperature \cite{Buescher05},
\begin{equation}
E^{D/N} = \frac{\hbar c}{2 \pi} \int_0^\infty dq_0
\, \mathrm{Tr} \log (M M_\infty^{-1}).
\label{eq:EDN}
\end{equation}
The total electromagnetic Casimir energy is the sum of the energies
$E^D$ and $E^N$.  The force between two objects separated by
$\vecseparation$ can be computed by differentiating the energy,
\begin{equation}
F^{D/N} = -\frac{\hbar c} {2 \pi} \int_0^\infty dq_0
\,
\mathrm{Tr} \,
M^{-1} \nabla_\vecseparation M.
\label{eq:FDN}
\end{equation}
The trace runs over the coordinates $\vecu$ and the indices $\alpha$,
$\beta$.  $M\inv_\infty$ is the functional inverse of $M$ with all
surfaces infinitely separated from one another.

Since the surfaces are static and since the system is invariant
under translations along the $u_1$-direction of the cylinder axes, it
is useful to transform the kernel $M$ to momentum space where it is
diagonal with respect to the momenta $q_0$ and $q_1$.  If we denote by
$\tilde M(q_0,q_1)$ the Fourier transformed matrix that is
non-diagonal with respect to the remaining momenta $q_2$, $q_2'$, the
energy can be expressed in terms of the determinant of that matrix,
\begin{equation}
\begin{split}
&E^{D/N} = \frac{\hbar c L} {8 \pi^2} \int dq_0 dq_1 \,
\log \|
\tM(q_0,q_1) \tMinfi(q_0,q_1) \|\\
&= \frac{\hbar c L} {4 \pi} \int_0^\infty q dq
\log 
\|\tM(q) \tMinfi(q)\|
\end{split}
\label{eq:EDN2}
\end{equation}
with $q=\sqrt{q_0^2+q_1^2}$ and $L$ the overall extent of the
system along the $u_1$-axis.  Hence \mtM is a block-diagonal matrix
of infinite size.  The blocks are indexed by $\alpha$, $\beta$ and the
momenta $q_2$, $q_2'$ which may take discrete or continuous values
depending on whether the corresponding surface is compact (cylinder)
or infinitely extended (plate) along the $u_2$-direction.  The matrix
elements are defined by the Fourier integrals
\begin{widetext}
\begin{equation}
\label{eq:D-def-elements}
\qtwobra \tM_{\alpha\beta}(q) \qtwopket
= \int G(\vecsa^\bot(u_2)-\vecsb^\bot(u_2');q)
e^{-iq_2 u_2 +iq_2' u_2'}
\frac{du_2 du_2'}{2\pi} 
\end{equation}
for D modes and 
\begin{equation}
\label{eq:N-def-elements}
\qtwobra \tM_{\alpha\beta}(q) \qtwopket 
 = \int
\partial_{\mathbf{n}_\alpha(u_2)} \partial_{\mathbf{n}_\beta(u'_2)}
G(\vecsa^\bot(u_2)-\vecsb^\bot(u_2');q)
e^{-iq_2 u_2 +iq_2' u_2'}
\frac{du_2 du_2'}{2\pi} 
\end{equation}
\end{widetext}
for N modes where $G(\vecx^\bot;q)$ is the $x_1$--Fourier
transformed free Green's function and $\vecsa^\bot(u_2)$ is the
projection of $\vecsa(u_2)$ onto the $x_2$-$x_3$-plane that is
perpendicular to the direction of translational invariance.

After obtaining the matrix elements the determinant of $\tM \tMinfi$
needs to be computed. For $n$ objects $\tM$ and $\tM \tMinfi$
obviously have $n\times n$ blocks each of which is indexed by $q_2$,
$q'_2$. For two objects we make use of the block matrix determinant
formula
\begin{eqnarray}
\label{eq:det-formula}
\left\|
\begin{pmatrix}
\tilde M_{11}&\tilde M_{12}\\ \tilde M_{21} &\tilde M_{22}
\end{pmatrix}
\right\| &=& \|\tilde M_{11}\|\|\tilde M_{22}-\tilde M_{21} \tilde M_{11}\inv \tilde M_{12}\| \nonumber\\
&=& \|\tilde M_{22}\| \|\tilde M_{11}-\tilde M_{12} \tilde M_{22}\inv \tilde M_{21} \| \, .
\end{eqnarray}
For more than two objects one could recursively reduce the size of the
matrix to compute its determinant. If the objects are infinite flat
plates, a simpler approach is to employ the method of images which
amounts to replace the free space Green's function by the Green's
function for half-spaces or slabs.

\subsection{Two cylinders}

\begin{figure}[h]
\centerline{ \epsfclipon \epsfysize=4.2cm
\epsfbox{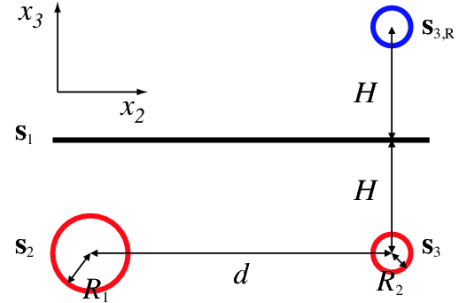} }
\caption{Surfaces and parameters used for computing matrix
  elements. Surface $S_3$ is displaced by $d$ to the right of
  surface $S_2$. Surface $S_3$ reflected at the plate yields surface $S_{3,R}$.}
\label{fig:surfaces}
\end{figure}

We begin by considering two cylinders at center-to-center distance $d$
of the general geometry shown in Fig.~\ref{fig:surfaces}. The
positions of the surfaces are parametrized in the $x_2$-$x_3$-plane
by
\begin{flalign}
\label{eq:s2-param}
\vecsperp_2(\phi) & = (R_1 \sin\phi , R_1 \cos\phi) \\
\label{eq:s3-param}
\vecsperp_3(\phi) & = (R_2 \sin\phi + d, R_2 \cos\phi) \, ,
\end{flalign}
where the parametrization coordinate here is $u_2=\phi$. With this
parametrization, the matrix elements defined in
Eqs.~\eqref{eq:D-def-elements}, \eqref{eq:N-def-elements} can be
computed straightforwardly. We find (for details see Appendix
\ref{app:matrix-elements})
\begin{flalign}
\label{eq:M22-D}
\mbra &\tM_{22} \mpket = \delta_{m,m'} I_m(R_1q) K_m(R_1q) \\
\label{eq:M23-D}
\mbra &\tM_{23} \mpket = (-i)^{m+m'} I_m(R_1q)I_{m'}(R_2q)K_{m-m'}(q d) .
\end{flalign} 
for D boundary conditions. The matrix elements of $\tM_{23}$ for N
boundary conditions are obtained through differentiation of the matrix
elements of $\tM_{23}$ for D boundary conditions with respect to $R_1$
and $R_2$. The elements of the diagonal blocks for N boundary conditions
are given by
\begin{equation}
\label{eq:M22-N}
\mbra \tM_{22} \mpket = \delta_{m,m'} q^2 I_m'(R_1q) K_m'(R_1q).
\end{equation}
The matrix elements of $\tM_{33}$ are given by Eqs.~\eqref{eq:M22-D}
and \eqref{eq:M22-N} with $R_1$ replaced by $R_2$ and
$\tM_{32}=\tM_{23}^\dagger$. Using the determinant formula
of Eq.~\eqref{eq:det-formula} and the fact that the off-diagonal
matrix elements vanish for $d\to\infty$, we get
\begin{equation}
\|\tM \tMinfi\| = \|\Imatrix - \tM_{32}\tM\inv_{22}\tM_{23}\tM\inv_{33}\| \, .
\label{eq:detMM2}
\end{equation}
Hence, the interaction energy of the two cylinders can be obtained
from Eq.~\eqref{eq:EDN2} and the matrix elements
\begin{equation}
\label{eq:2-cyl-elements}
\begin{split}
&\mbra \tM_{32}\tM\inv_{22}\tM_{23}\tM\inv_{33} \mpket \\
& = 
\ZX(1,1)
\sum_{n} K_{m+n}(qd) Z^X_{n,n}(2,2) K_{n+m'}(qd) \, ,
\end{split}
\end{equation}
with $X=D$, $N$ standing for Dirichlet and Neumann boundary conditions
and the definitions
\begin{eqnarray}
Z^D_{m,m'}(i,j) &=& \frac{I_m(R_i q)}{K_{m'}(R_j q)} \\
Z^N_{m,m'}(i,j) &=& \frac{I_m'(R_i q)}{K_{m'}'(R_j q)} \, .
\end{eqnarray}

\subsection{Cylinders and plates: Method of images}

Now we add infinite plates to the geometry, see
Fig.~\ref{fig:surfaces}. The effective number of surfaces can be kept
the same when adding the plates to the system if instead of the free
space Green's function, $G_0$, modified Green's functions are used,
which obey Dirichlet or Neumann boundary conditions at the plates.
For one plate the half-space Green's function is
\begin{equation}
G\onep^X(\vecx,\vecx';q_0) = G_0(\vecx,\vecx';q_0)-\sX G_0(\vecx,\vecx'_R;q_0)
\label{eq:Gonep}
\end{equation}
where $\vecx'_R=(x_1,x_2,-x_3+2H)$ is the reflection of $\vecx'$ at
the plane at $x_3=H$ and $s^D=+1$, $s^N=-1$. For two plates infinitely
many images have to be used since successive reflections at the plates
generate a series of images with increasing separation from the
plates. For this slab geometry, the Green's function can be written as
the series
\begin{equation}
\label{eq:Gtwop}
\begin{split}
G\twop^X(\vecx&,\vecx';q_0) = G_0(\vecx,\vecx';q_0)\\
 - \sX &[G_0(\vecx,\vecx'_{R1};q_0) + G_0(\vecx,\vecx'_{R2};q_0)] \\
 + \quad\, &[G_0(\vecx,\vecx'_{R1,R2};q_0) + G_0(\vecx,\vecx'_{R2,R1};q_0)] \\
 - \sX &[G_0(\vecx,\vecx'_{R1,R2,R1};q_0) + G_0(\vecx,\vecx'_{R2,R1,R2};q_0)] \\
 + \quad\,  &\cdots \, ,
\end{split}
\end{equation}
where $\vecx'_{R\alpha,R\beta,...}$ is obtained from $\vecx'$ by a
sequence of reflections at plate $\alpha$, $\beta$, $\ldots$ The
theory developed for the free space at the beginning of this Section
applies also to half-space and slab geometry. The interaction between
a set of cylinders in the presence of one or two parallel plates can
be obtained again from Eq.~\eqref{eq:EDN2}; one only needs to
substitute the half-space or slab Green's functions for $G$ in
Eqs.~\eqref{eq:D-def-elements}, \eqref{eq:N-def-elements} and the obtain
cylinder-cylinder matrix elements. When using these image Green's
functions, their appropriate Fourier transformed matrices shall be
labeled $\tMonep$ and $\tMtwop$.

Below, we shall consider the two cylinders $S_2$ and $S_3$, see
Fig.~\ref{fig:surfaces}.  The surfaces of the reflected cylinders
have the parametrization
\begin{flalign}
\label{eq:s2R-param}
\vecsperp_{2,R}(\phi) & = (R_1 \sin\phi , \pm R_1 \cos\phi + \delta) \\
\label{eq:s3R-param}
\vecsperp_{3,R}(\phi) & = (R_2 \sin\phi + d, \pm R_2 \cos\phi + \delta)\, ,
\end{flalign}
where $\delta$ is a distance normal to the plates.  Upon one
reflection $\delta$ is $2H$ and the cylinder surface orientation
is reversed, hence, the minus sign is chosen in
Eqs.~\ref{eq:s2R-param} and \ref{eq:s3R-param}. For two plates the
reflected cylinder is reflected again, so that the plus sign must be
chosen for the orientation, and $\delta=-2H_1-2H_2$ (if the first
plate is loacted a distance $H_1$ above and the second plate a
distance $H_2$ below the cylinder, see
Fig.~\ref{fig:cyl2plate12surf}).

The corresponding matrix elements of Eqs.~\eqref{eq:D-def-elements},
\eqref{eq:N-def-elements} for D modes are
\begin{flalign}
\label{eq:M2-2R-D}
\mbra &\tM_{2(2,R)} \mpket = (\mp 1)^{m'}
I_m(R_1q)I_{m'}(R_1q)K_{m\mp m'}(q \delta) \notag.\\
\mbra &\tM_{2(3,R)} \mpket = (\mp 1)^{m'}
\left(\frac{\delta-id}{\sqrt{\delta^2+d^2}}\right)^{m\mp m'}\\
&\times I_m(R_1q)I_{m'}(R_2q)K_{m\mp m'}(q\sqrt{\delta^2+d^2}) \notag.
\end{flalign}
The elements of $\tM_{3(3,R)}$ follow from those of $\tM_{2(2,R)}$ by
replacing $R_1$ by $R_2$ and the elements of $\tM_{3(2,R)}$ are given
by those of $\tM_{2(3,R)}$ with $R_1$ and $R_2$ interchanged and $d$
replaced by $-d$.  For N modes, the elements of $\tM_{2(3,R)}$ are
obtained by differentiation of the D elements with respect to $R_1$
and $R_2$. The elements of $\tM_{2(2,R)}$ for N modes are given by 
\begin{equation}
  \label{eq:M2-2R-N}
  \mbra \tM_{2(2,R)} \mpket = (\mp 1)^{m'} q^2
I'_m(R_1q)I'_{m'}(R_1q)K_{m\mp m'}(q \delta) \, .
\end{equation}

\section{Interaction energies}
\label{sec:int_energies}



\subsection{Two cylinders}

\begin{figure}[h]
\centerline{ \epsfclipon \epsfysize=2cm
\epsfbox{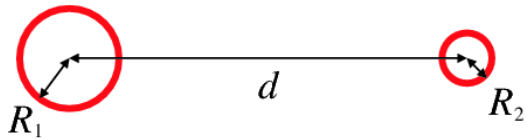} }
\caption{Cylinder-cylinder geometry.}
\label{fig:cylcylsurf}
\end{figure}

We consider two cylinders of radii $R_1$ and $R_2$ with
center-to-center separation $d$, see Fig.~\ref{fig:cylcylsurf}. For
this geometry the interaction energy is obtained from
Eqs.~(\ref{eq:EDN2}) and (\ref{eq:detMM2}) with the matrix elements of
Eq.~(\ref{eq:2-cyl-elements}). For large separations $d\gg R_1,\,R_2$,
the asymptotic behavior of the energy is determined by the matrix
elements for $m=m'=0$ for D modes and $m=m'=0,\, \pm 1$ for N
modes. Taking the determinant of the matrix that consists only of
these matrix elements and integrating over $q$ yields
straightforwardly the attractive interaction energies
\begin{equation}
\label{eq:E-two-cylinders}
\begin{split}
E^D & = -\frac{\hbar c L}{d^2}\frac{1}{8\pi\log(d/R_1)\log(d/R_2)} \\
& \times \left(1-\frac{1}{\log(d/R_1)}-\frac{1}{\log(d/R_2)} + \ldots \right) \, , \\
E^N & = -\hbar c L \frac{7}{5\pi}\frac{R_1^2 R_2^2}{d^6} \, .
\end{split}
\end{equation}
The asymptotic interaction is dominated by the contribution from TM (D) modes
that vanishes for $R_\alpha\to 0$ only logarithmically.

\begin{figure}[h]
\centerline{ \epsfclipon \epsfxsize=\linewidth 
\epsfbox{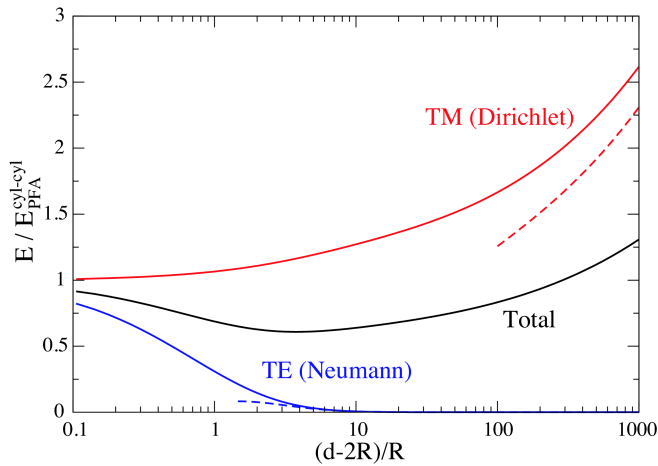} }
\caption{Casimir energy for two cylinders of equal radius $R$ as a
  function of surface-to-surface distance $d-2R$ (normalized by the
  radius). The energy is divided by the PFA estimate $\EPFAcylcyl$ for
  the energy given in the introduction. The solid curves show our
  numerical results; the dashed lines represent the asymptotic results
  of Eqs.~\ref{eq:E-two-cylinders}. The $1/\log$ corrections to the
  leading order result for TM modes cause very slow convergence.}
\label{fig:cyl2energy}
\end{figure}

For arbitrary separations higher order partial waves have to be
considered. The number of partial waves has to be increased with
decreasing separation. A numerical evaluation of the determinant and
the $q$-integration can be performed easily and reveals an
exponentially fast convergence of the energy in the truncation order
for the partial waves.  Down to small surface-to-surface separations
of $(d-2R)/R=0.1$ we find that $m=40$ partial waves are sufficient to
obtain precise results for the energy. The corresponding result for
the energies of two cylinders of equal radius is shown in
Fig.~\ref{fig:cyl2energy}. Notice that the minimum in the curve for
the total electromagnetic energy results from the scaling by the PFA
estimate of the energy. The total energy is monotonic and the force
attractive at all separations.

\subsection{Cylinder-plate geometry}

\begin{figure}[h]
\centerline{ \epsfclipon \epsfysize=2cm
\epsfbox{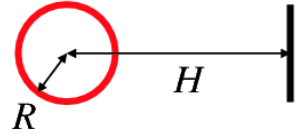} }
\caption{Cylinder-plate geometry.}
\label{fig:cylplatesurf}
\end{figure}

\begin{figure}[h]
\centerline{ \epsfclipon \epsfxsize=\linewidth 
\epsfbox{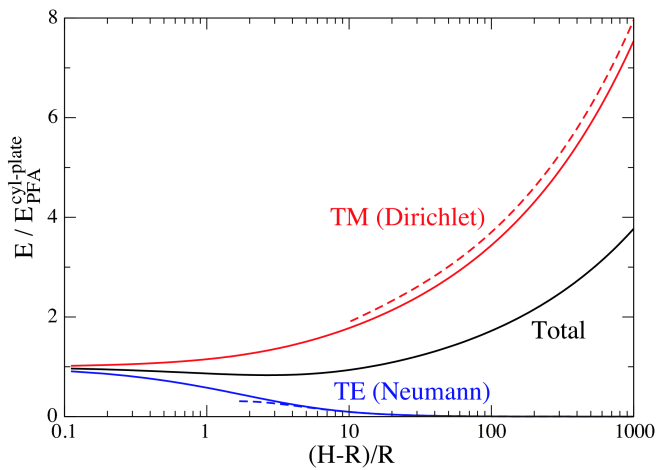} }
\caption{Casimir energy for one cylinder of radius $R$ parallel to one
  plate as a function of the surface-to-surface distance $H-R$
  (normalized by the radius). The energy is divided by the PFA
  estimate $\EPFAcylplate$ of energy given in the Introduction. The
  solid curves reflect our numerical results; the dashed lines
  represent the asymptotic results of
  Eqs. \ref{eq:E-cyl-plate}. Convergence for the TM (Dirichlet) energy
  to the asymptotic result is very slow because of $1/\log$
  corrections.}
\label{fig:cyl1plate1energy}
\end{figure}

The simplest geometry to which the method of images can be applied is
composed of an infinite plate and a parallel cylinder of radius $R$,
see Fig.~\ref{fig:cylplatesurf}. The Casimir energy for this geometry
has been computed in Ref.~\cite{Emig06} from the matrix elements of
Eqs.~(\ref{eq:D-def-elements}), (\ref{eq:N-def-elements}) for the
plate and the cylinder. Here, we employ the method of images so that
we have to consider only one surface, the cylinder, that is placed
into the half-space that is bounded by the plate. Hence, we substitute
$G$ by $G_{1p}$ in Eq.~(\ref{eq:D-def-elements}),
(\ref{eq:N-def-elements}) together with the parametrizations of
Eq.~(\ref{eq:s3-param}) and of Eq.~(\ref{eq:s3R-param}) with the $+$
sign, $R_2=R$ and $\delta=2H$.  The resulting $\tMonep$ is simply one
block for the cylinder since there is effectively only one surface,
and it equals $\tM_{33}-\sX\tM_{3(3,R)}$. When the plate is moved to
infinite separation from the cylinder, there is only the free
cylinder, so $\tMonepinfi=\tM\inv_{33}$. The determinant of
Eq.~(\ref{eq:EDN2}) can now be written as
\begin{equation}
\|\tM_{1p} \tM_{1p,\infty} \| = \|\Imatrix -
\sX\tM_{3(3R)}\tM\inv_{33} \| \, .
\label{eq:det-cyl-plate}
\end{equation}
The matrix elements of $\tM_{3(3R)}\tM\inv_{33}$ are given for D modes
by
\begin{equation}
  \label{eq:plate-cyl-elements-D}
  \mbra \tM_{3(3R)}\tM\inv_{33} \mpket = \frac{I_m(Rq)}{K_{m'}(Rq)} K_{m+m'}(2Hq)   
\end{equation}
and for N modes by
\begin{equation}
  \label{eq:plate-cyl-elements-N}
  \mbra \tM_{3(3R)}\tM\inv_{33} \mpket =  \frac{I'_m(Rq)}{K'_{m'}(Rq)} K_{m+m'}(2Hq) \, .
\end{equation}
This result in combination with Eq.~(\ref{eq:EDN2}) is identical to
the one given in Eqs.~(5)-(8) of Ref.~\cite{Emig06}. The asymptotic
expression for the attractive interaction energy at $H\gg R$ reads
\begin{equation}
\begin{split}
E^D & = -\frac{\hbar c L}{H^2}\frac{1}{16\pi\log(H/R)}, \\
E^N & = - \hbar c L \frac{5}{32\pi} \frac{R^2}{H^4} \, .
\end{split}
\label{eq:E-cyl-plate}
\end{equation}
The total electromagnetic Casimir interaction is again dominated by
the contribution from the D mode with $m=0$ which depends only
logarithmically on the cylinder radius.  The interaction at all
separations follows, as in the case of two cylinders, from a numerical
computation of the determinant of Eq.~(\ref{eq:det-cyl-plate}) and
integration over $q$.  The result is shown in
Fig.~\ref{fig:cyl1plate1energy}.

\subsection{Two cylinders, parallel to plate(s)}

\begin{figure}[h]
\centerline{ \epsfclipon \epsfysize=4.2cm
\epsfbox{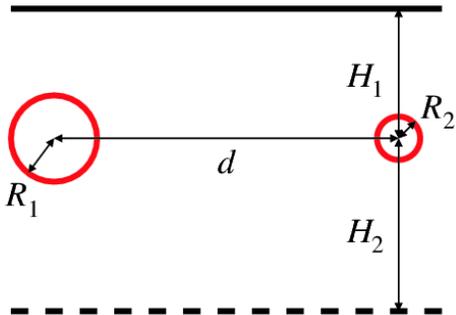} }
\caption{Two cylinders parallel to one plate or sandwiched between two parallel plates.}
\label{fig:cyl2plate12surf}
\end{figure}

The image technique lends itself to studying multibody interactions
involving plates and cylinders. We consider the geometry shown in
Fig.~\ref{fig:cyl2plate12surf} with two cylinders that are placed
parallel to one or in-between two parallel plates. Rodriguez et
al. \cite{Rodriguez07:PRL} studied a similar geometry consisting of
two metal squares between two parallel metal sidewalls by computing
numerically the mean stress tensor and observed that the force between
the two squares changes non-monotonically when the two plates are
pulled away. In previous work, we applied the stress tensor method and
the path integral approach presented above to the geometry of
Fig.~\ref{fig:cyl2plate12surf} and found again a non-monotonic
dependence of the force between the cylinders on the separation
between the plates \cite{Rahi:2008mz}. Here we provide the technical
details of the path integral approach and the method of images
employed in the latter work.

\begin{figure}[ht]
\centerline{ \epsfclipon \epsfxsize=\linewidth 
\epsfbox{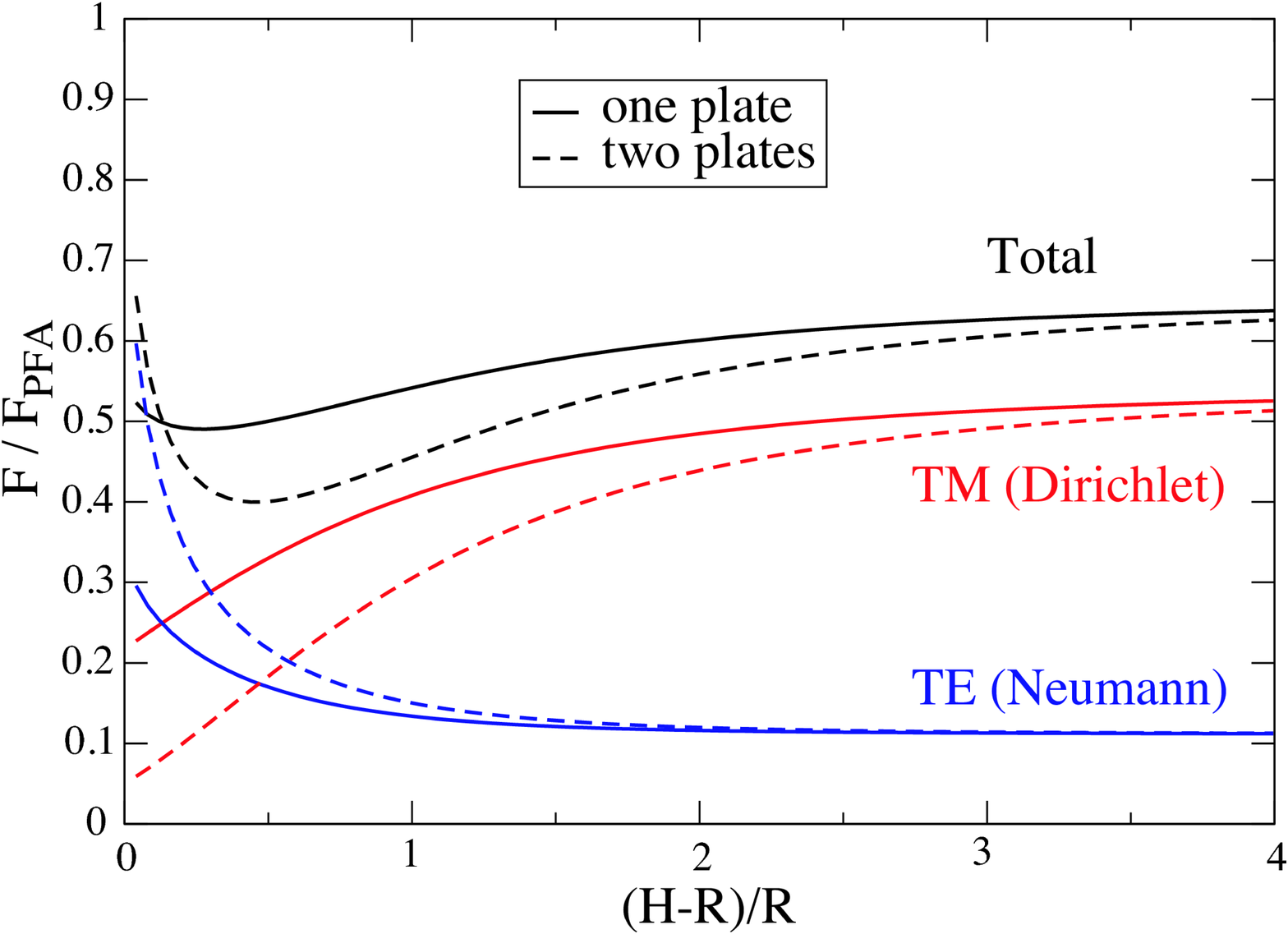} }
\caption{Electromagnetic Casimir force per unit length between two
  cylinders for the geometry of Fig.~\ref{fig:cyl2plate12surf} with
  $H_1=H_2=H$ and $R_1=R_2=R$ vs.  the ratio of sidewall separation to
  cylinder radius $(H-R)/R$, at fixed distance $(d-2R)/R=2$ between
  the cylinders, normalized by the total PFA force per unit length
  between two isolated cylinders [$F_{\text{PFA}}= \frac{5}{2}(\hbar
  c\pi^3/ 1920)\sqrt{R/(d-2R)^7}$ \cite{Rahi:2008mz}]. The force is
  attractive. The solid lines refer to the case with one sidewall,
  while dashed lines depict the results for two sidewalls.  Also shown
  are the individual TE (blue) and TM (red) forces.}
\label{fig:forcecylcyl}
\end{figure}

To study the interaction between the cylinders and plates it turned
out to be more convenient to compute directly the forces between the
objects as defined by Eq.~\eqref{eq:FDN}. For the force between the
two cylinders we set $\vecseparation = d \hat {\bf x}_2$ and for the
force between the cylinders and a plate we choose $\vecseparation =
H_1 \hat {\bf x}_3$. The matrix $M$ of Eq.~\eqref{eq:FDN} can be
constructed by the method of images as follows.  We consider the two
cylinders as the surfaces that are placed either inside a half-space
or a slab so that the matrix elements of $M$ are given by
Eqs.~\eqref{eq:D-def-elements}, \eqref{eq:N-def-elements} with $G$
replaced by $G^X_{1p}$ or $G^X_{2p}$ of Eqs.~\eqref{eq:Gonep},
\eqref{eq:Gtwop}, respectively.  We shall refer to the corresponding
$2\times 2$ block matrices $\tMonep$ and $\tMtwop$ matrices as
$\tMnp$ which is of the form
\begin{equation}
  \label{eq:M_np}
  \tMnp =
\begin{pmatrix}
\tM_{\text{np},22}& \tM_{\text{np},23} \\
\tM_{\text{np},32}& \tM_{\text{np},33}
\end{pmatrix}
 \, .
\end{equation}
In the two-body cylinder case without sidewalls the
$\tM_{\alpha\alpha}$ matrix blocks describing the self-interaction
were diagonal in $m$. Here, the self-interaction blocks
$\tM_{\text{np},\alpha\alpha}$ contain image information and are not
diagonal. The matrix elements of $\tMnp$ are constructed from the
elements of Eqs.~\eqref{eq:M2-2R-D}, \eqref{eq:M2-2R-N}.  Their
explicit form and a formula for the inversion of $\tMnp$ are given in
Appendix \ref{app:multibody}. The integrand of Eq.~(\ref{eq:FDN}) can
be straightforwardly computed by truncating the matrix $M$ at a finite
partial wave order $m$ and performing the matrix multiplication and
trace in Fourier space. Including up to $m=35$ partial waves, we
obtain for the Casimir force between two cylinders of equal radii in
the presence of one or two sidewalls the results shown in
Fig.~\ref{fig:forcecylcyl}. In this figure the force at a fixed
surface-to-surface distance $d-2R=2R$ between the cylinders is plotted
as a function of the relative separation $(H-R)/R$ between the plate
and cylinder surfaces. Two interesting features can be
observed. First, the attractive total force varies non-monotonically
with $H$: Decreasing for small $H$ and then increasing towards the
asymptotic limit between two isolated cylinders for large $H$,
cf. Eq.~(\ref{eq:E-two-cylinders}). The extremum for the one-sidewall
case occurs at $H-R \approx 0.27 R$, and for the two-sidewall case is
at $H-R \approx 0.46 R$. Second, the total force for the two-sidewall
case in the proximity limit $H=R$ is larger than for $H/R
\rightarrow\infty$. As might be expected, the $H$-dependence for one
sidewall is weaker than for two sidewalls, and the effects of the two
sidewalls are not additive: not only is the difference from the
$H\rightarrow\infty$ force not doubled for two sidewalls compared to
one, but the two curves actually intersect at a separation of
$H/R=1.13$.

A simple generic argument for the non-monotonic sidewall effect has
been given in Ref.~\onlinecite{Rahi:2008mz}. It arises from a
competition between the force from TE and TM modes as demonstrated by
the results in Fig.~\ref{fig:forcecylcyl}. An intuitive perspective
for the qualitatively different behavior of the TE and TM force as a
function of the sidewall distance is obtained from the method of
images.  For the D modes (TM polarization) the Green's function of
Eq.~(\ref{eq:Gonep}) is obtained by subtracting the contribution from
the image so that the image sources have \emph{opposite} signs.  Any
configuration of fluctuating TM charges on one cylinder is thus
screened by images, more so as $H$ is decreased, \emph{reducing} the
force on the fluctuating charges of the second cylinder.  This is
similar to the effect of a nearby grounded plate on the force between
two opposite electrostatic charges. Since the reduction in force is
present for every charge configuration, it is there also for the
average over all configurations, accounting for the variations of the
TM curves in Fig.~\ref{fig:forcecylcyl}.

By contrast, the N modes (TE polarization) require image sources of
the \emph{same} sign as demonstrated by the half-space Green's
function of Eq.~(\ref{eq:Gonep}).  The total force between fluctuating
sources on the cylinders is now larger and increases as the plate
separation $H$ is reduced.  (An analogous additive effect occurs for
the classical force between current loops near a conducting plane.)
Note, however, that while for each fluctuating source configuration,
the effect of images is additive, this is not the case for the
average over all configurations. More precisely, the effect of an
image source on the Green's function is not additive because of
feedback effects: the image currents change the surface current
distribution, which changes the image, and so forth.  For example, the
net effect of the plate on the Casimir TE force \emph{is not} to
double the force as $H\to R$.  The increase is in fact
larger than two due to the correlated fluctuations.

\begin{figure}[h]
\centerline{ \epsfclipon \epsfxsize=\linewidth 
\epsfbox{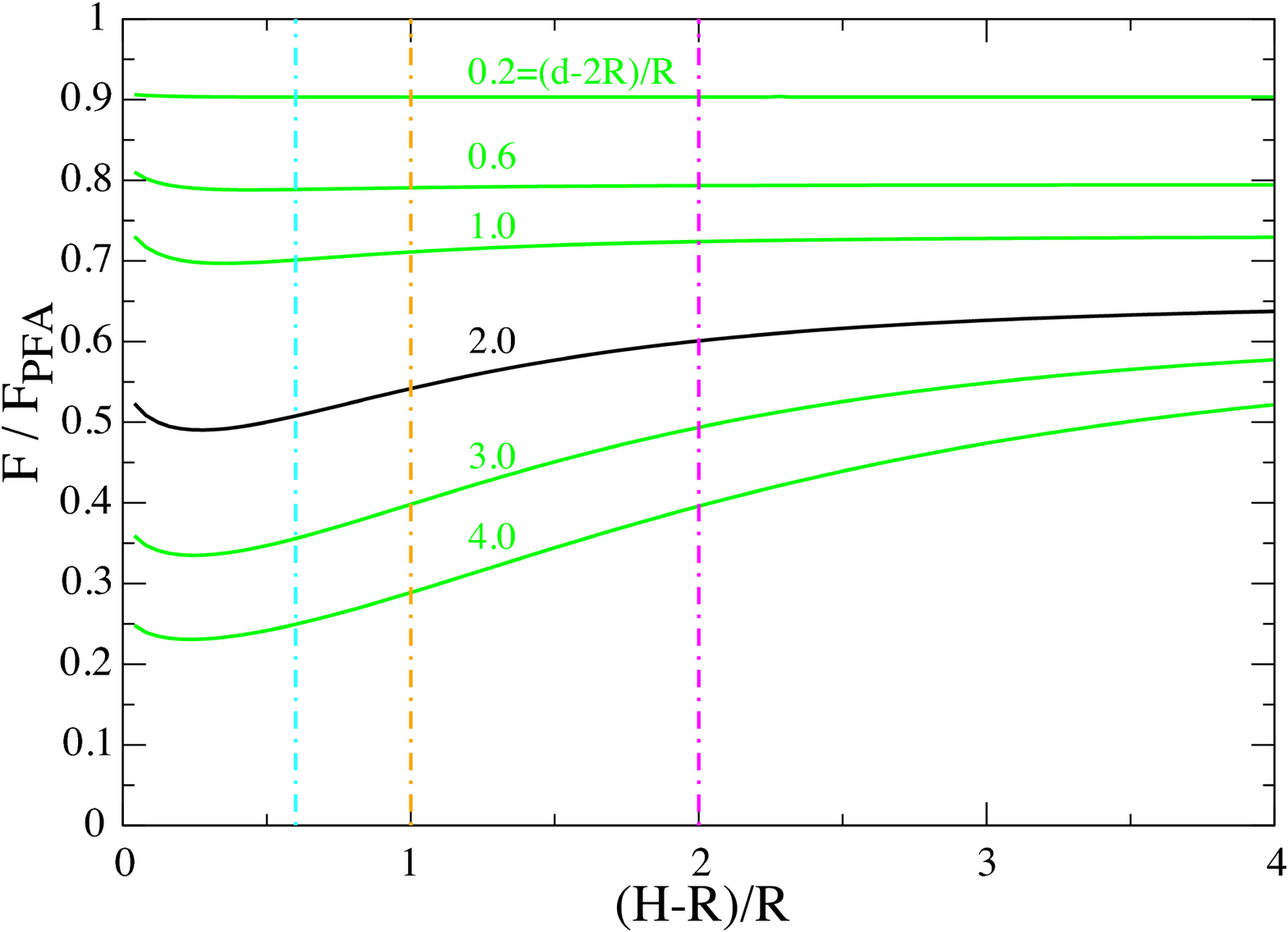} }
\caption{Casimir force per unit length between two cylinders of equal
  radius vs. the ratio of sidewall separation to cylinder radius
  $(H-R)/R$ (for one plate), normalized by the total PFA force per
  unit length between two isolated cylinders for various
  $(d-2R)/R$. The non-monotonic effect appears to become weaker as
  $(d-2R)/R$ is moved away from 2. The vertical lines indicate the
  sidewall separations used in Fig.~\ref{fig:forcecylplate}.}
\label{fig:forcecylcylmany}
\end{figure}

In Fig.~\ref{fig:forcecylcylmany}, we show the total force between the
cylinders vs. the sidewall separation $H/R$ for a variety of different
values of the cylinder separation $d/R$ in the presence of a single
sidewall. As we vary $d/R$ the depth of the minimum in the force
changes, see Fig.~\ref{fig:forcecylcylmany}. The separation $(d-2R)/R
= 2$ from Fig.~\ref{fig:forcecylcyl} seems to achieve the largest
value of non-monotonicity. For larger or smaller $d$ the degree of
non-monotonicity (defined as the difference between the minimum force
and the force in the limit $H\to 0$) decreases.  For small $d$, the
force approaches to PFA estimate. For large $d$, the TM (Dirichlet)
force dominates except when the cylinders are sufficiently close to
the metal plate when it is reduced enough by its image cylinder that
the TE (Neumann) force takes over.

\begin{figure}[h]
\centerline{ \epsfclipon \epsfxsize=\linewidth
\epsfbox{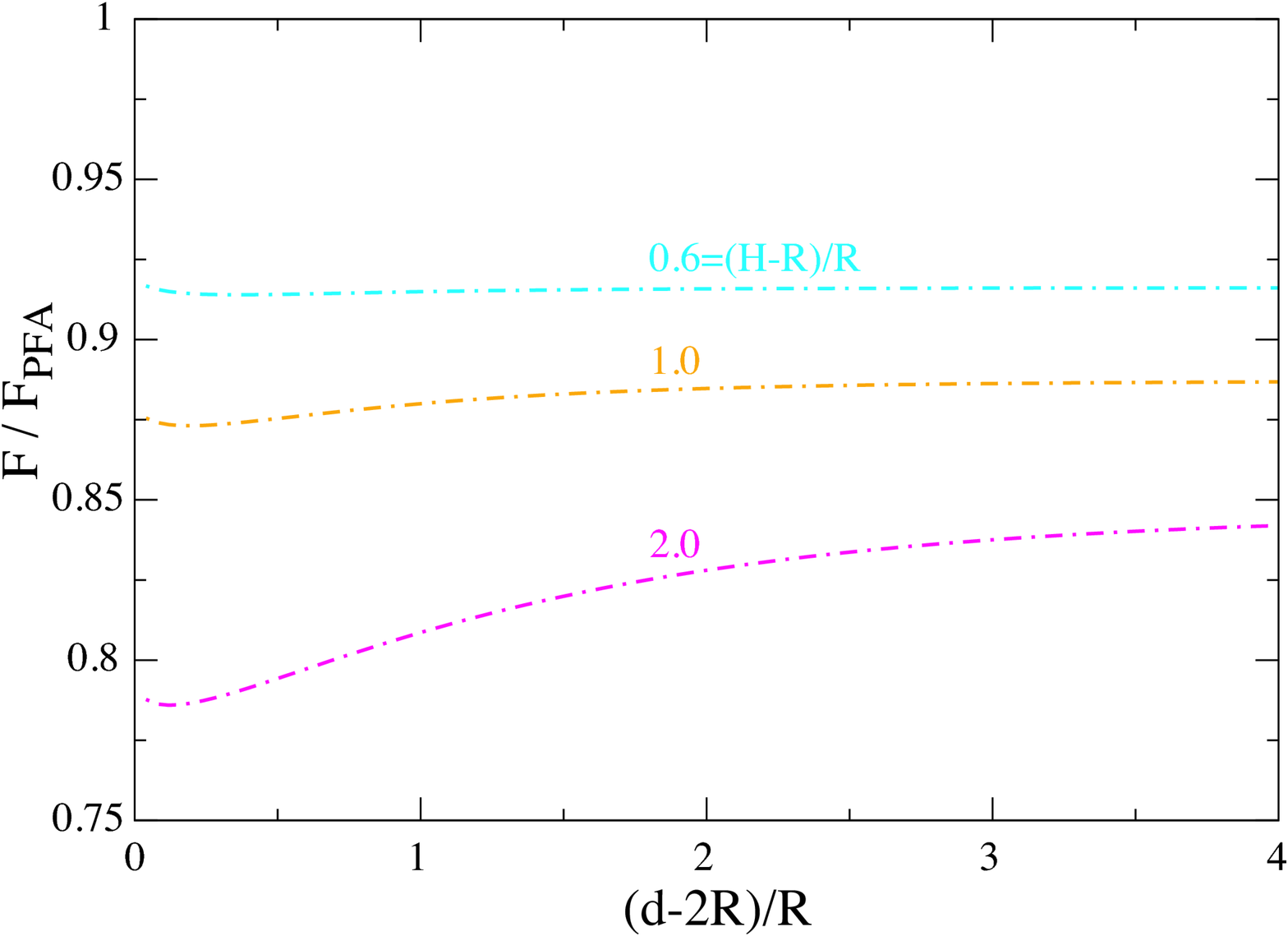} }
\caption{Total Casimir force between two cylinders of equal radius $R$
    and a sidewall vs.~the ratio of cylinder surface separation to
    cylinder radius $(d-2R)/R$, normalized by the total PFA force per
    unit length between a cylinder and a plate $F_{\text{PFA}}=
    \frac{5}{2}(\hbar c\pi^3/ 960)\sqrt{R/2(d-2R)^7}$ \cite{Dalvit04}
    for plate separations of $H-R = 0.6 R$, $R$, and $2R$.  Note that
    the normalization is different from the cylinder-cylinder PFA in
    the previous figures.}
\label{fig:forcecylplate}
\end{figure}

While the above arguments explain the competition between TE and
TM forces, they do not show that the sum of these competing forces is
nonmonotonic. For example, if the TE and TM variations with $H$ were
equal and opposite, they would cancel with no net dependence on $H$.
That this is not the case can be checked by examining the two
limits $H-R\gg R$ and $H-R\ll R$. In order to simplify the analysis,
we assume that there is one sidewall and that in both limits the two
cylinders have a large separation $d\gg H-R$, $R$. Then in the case
$H-R\gg R$ the cylinders and their images are separated by a distance
that is large compared to $R$ so that the forces are dominated by the
lowest partial waves, $s$-wave for TM and both $s$- and $p$-wave for
TE modes \cite{Emig06}. The former is stronger and dominates the
asymptotic force for which we obtain
\begin{equation}\label{asymptote}
\frac{F_D}{L} = - \frac{4 \hbar c}{\pi}\frac{  H^4}{d^7 \ln^2(R/H)},
\end{equation}
confirming the reduced net force as the cylinders get closer to the
plate.

It is instructive to justify the scaling of the force in
Eq. (\ref{asymptote}) with $H$ and $d$ from simple physical arguments.
While the logarithmic dependence on $R$ could have been
anticipated~\cite{Emig06}, the $H^4$ scaling is a remarkable
consequence of the multi-body effect. For TM modes, the field obeys D
boundary conditions so that the mirror source has opposite
sign. Therefore, each cylinder and its mirror image can be considered
as a dipole of size $\sim H$.  The interaction of the two dipoles
should scale as the interaction between two cylinders of size $\sim H$
with Neumann boundary conditions.  For $d\gg H$ the force for the
latter problem scales as $\sim H^4/d^7$, explaining the above
result~\cite{Emig06}, up to the logarithm.

In the opposite limit $H-R \ll R$, the image cylinder is very close to
the original cylinder so that the interaction involves partial waves
of high order. In an attempt to analytically understand this limit, we
performed certain asymptotic calculations reported in Appendix
\ref{app:asymp}. Since the relevance of the conclusions to the regime
studied numerically (Fig. \ref{fig:forcecylcyl}) is not clear, we
shall not further explore the implications of these results. We note,
however, that the numerical results in the regime of
Fig. \ref{fig:forcecylcyl} support the dominance and fast decay of TE
modes for this range of parameters, as justfification for the
non-monotonic behavior.

Thus far we have discussed the variation of the force between
the two cylinders with the sidewall separation. We found that
the force is not monotonic in $H$. This also implies that the force
between the cylinders and the sidewalls is not monotonic in $d$.
A nonmonotonic force $F_{x_2}$ between the cylinders means that
there is a value of $H$ where $\partial F_{x_2}/\partial H = 0$.
Since the force is the derivative of the energy, $F_{x_2} =
-\partial\mathcal{E}/\partial d$, at this point
$\partial^2\mathcal{E}/\partial d \partial H = 0$.  These two
derivatives, of course, can be interchanged to yield
$\partial(\partial\mathcal{E}/\partial H)/\partial d = 0$.  But this
means that $\partial F_{x_3}/\partial d = 0$ at the same point,
where $F_{x_3} = -\partial\mathcal{E}/\partial H$ is the force between
the cylinders and the sidewall.  This cylinders-sidewall force is
plotted in Fig.~\ref{fig:forcecylplate} as a function of $d/R$ for
various values of $H/R$ and clearly is non-monotonic in $d/R$. The
non-monotonicity is smaller which is not surprising since the effect
of a small cylinder on the force between two bodies is smaller than
the effect of an infinite plate.


\section{Discussion}
\label{sec:discussion}

In previous research, unusual Casimir force phenomena were sought by
considering parallel plates with exotic materials: for example,
repulsive forces were predicted using magnetic
conductors~\cite{Kenneth02}, combinations of different
dielectrics~\cite{Imry05}, fluids between the plates~\cite{Munday07},
and even negative-index media with gain~\cite{Leonhardt07}.  A
different approach is to use idealized materials such as perfect
conductors with more complicated geometries: as illustrated in this
and previous~\cite{Rodriguez07:PRL} work, surprising nonmonotonic
(attractive) effects can arise by considering as few as three
objects. These effects arise from the collective properties of
fluctuation forces and cannot emerge in a system of particles that
interact by a pairwise two-body potential.

It would be interesting to probe the collective nature of fluctuation
forces in experiments. So far, only the interaction between two
objects, mostly for sphere-plate geometries, have been realized
experimentally. Among the forces studied here, the one between two
cylinders and a plate, see Fig.~\ref{fig:forcecylplate}, might be most
feasible in experimental studies.  To measure the cylinder-plate
force, the two cylinders need not be separated by vacuum---we expect
that a similar phenomenon will arise if the cylinders are separated by
a dielectric spacer layer of fixed thickness. This avoids the problem
of parallelism one would face when measuring the force between two
cylinders.  Unfortunately, the nonmonotonic effect in
Fig.~\ref{fig:forcecylplate} is rather small (roughly 0.2\%), but it
may be possible to increase it by further optimization of the geometry
and/or the material.

Another important issue is the translational invariance of the
geometries considered here. The geometries of experimental tests will
obviously lack this symmetry beyond some length scale. Hence it is
important to study deviations from the here considered quasi-2D
geometries due to cylinders and plates of finite size. Then the
surfaces have to be treated as compact objects. For this full
3-dimensional problem, TM and TE modes are longer decoupled so that
the full electromagnetic vector field has to be quantized. This can be
done by a recently developed multipole expansion which yields the
Casimir interaction between objects of arbitrary shape in terms of
their T-matrices \cite{Emig07,Emig08}.  At asymptotically large
separations between the objects, the interaction is determined to
leading order by the object's static electric and magnetic dipole
polarizabilities.  For a cylinder of finite length $L \gg R$, the
component of the electric polarizability tensor along the cylinder
axis scales as $L^3/\log(L/R)$. Hence, the interaction energy between
two parallel cylinders of finite length behaves for separations $d\gg
L \gg R$ as $\sim -\hbar c [L^6/\log^2(L/R)]d^{-7}$. This result shows
that for large $d$ the interaction is no longer proportional to $L$ or
the product of the volume of the cylinders as might be expected in
analogy with the Casimir-Polder interaction between spherical
particles \cite{Emig06}. In fact, for large $d$ the interaction
amplitude scales as the product of the cubes of the
largest dimensions of the objects.  The cross-over between
the two extreme cases of infinitely long cylinders and asymptotically
large separations between finite-sized cylinders can be obtained in
principle from a multipole expansion by including higher order
multipoles.  Finally, we note that all results given for infinitely
long cylinders here can be also obtained within the T-matrix approach
\cite{Emig07,Emig08}. The latter technique should be particularly
useful to determine whether the same non-monotonic effects occur for
two spheres next to a metal plate.

\begin{acknowledgments}
  This work was supported in part by NSF grant DMR-04-26677 (SJR and
  MK), by the US Dept.~of Energy (DOE) under cooperative research
  agreement DF-FC02-94ER40818 (RLJ), and by DFG grant EM70/3 (TE).
\end{acknowledgments}


\begin{appendix}

\section{Matrix elements}
\label{app:matrix-elements}

Here, we derive $\tM_{22}$ and $\tM_{23}$ based on
Eqs.~\eqref{eq:D-def-elements} and \eqref{eq:N-def-elements} for TM
(Dirichlet) and TE (Neumann) modes in detail.

We compute $\tM_{22}$ for Dirichlet boundary conditions as follows:
\begin{equation}
\begin{split}
& \mbra \tM_{22} \mpket = \\
& \int G(\vecs_2^\bot(\phi)-\vecs_2^\bot(\phi');q)
e^{-i m \phi +i m' \phi'}
\frac{d\phi d\phi'}{2\pi} = \\
& \int \sum_j e^{ij(\phi-\phi')} I_j(q r_<) K_j(q r_>)
e^{-i m \phi +i m' \phi'}
\frac{d\phi d\phi'}{(2\pi)^2} = \\
& \delta_{m,m'} I_m(R_1q) K_m(R_1q).
\end{split}
\end{equation}
The Green's function expansion in terms of modified Bessel functions
of the first ($I$) and second kind ($K$) contains arguments $r_<$ and
$r_>$, for which the radius of $\vecs_2^\bot$ is inserted. Once the
angular integrations are carried out one obtains in the last line the
result of Eq.~\eqref{eq:M22-D}.

In order to compute $\tM_{22}$ for Neumann boundary conditions with
the least effort, we make use of the previous calculation. According
to Eq.~\eqref{eq:N-def-elements} the directional derivatives with
respect to the surface normal of cylinders $\vecs_2^\bot(\phi)$ and
$\vecs_2^\bot(\phi')$ need to be taken inside the integral. We simply
assume that $\vecs_2^\bot(\phi)$ and $\vecs_2^\bot(\phi')$ are two
different concentric cylinders with different radii $R_1$ and $R_1'$,
respectively, and $R_1<R_1'$ so that the derivatives
$\partial_{\mathbf{n}_2(\phi)}
\partial_{\mathbf{n}_2(\phi')}$ are taken as $\partial_{R_1}
\partial_{R_1'} I_j(q R_1) K_j(q R_1')$. Then, the limit $R_1' \to
R_1$ is taken to yield Eq. \eqref{eq:M22-N}.

The computation of $\tM_{23}$ is more involved and is carried out
without a convenient expansion of the Green's
function. Eqs.~\eqref{eq:s2-param} and \eqref{eq:s3-param} give the
surface parametrizations which yield
\begin{equation}
\begin{split}
& \mbra \tM_{23} \mpket = \\
& \int G(\vecs_2^\bot(\phi)-\vecs_3^\bot(\phi');q)
e^{-i m \phi +i m' \phi'}
\frac{d\phi d\phi'}{2\pi} = \\
&\int
\left[
\frac{e^{i \veck (\vecs_2^\bot(\phi)-\vecs_3^\bot(\phi')}e^{-im\phi +im' \phi'}}
{k^2+q^2}
\right]
\frac{d^2k d\phi d\phi'}{(2\pi)^3} = \\
&\int
\frac{e^{i (k_1 R_1 \sin\phi + k_2 R_1 \cos\phi - k_1 (R_2 \sin\phi' +d)- k_2 R_2 \cos\phi')}}{k^2+q^2} \\
&\qquad\times e^{-im\phi +im' \phi'}
\frac{d^2k d\phi d\phi'}{(2\pi)^3} = \\
&\int
\frac{e^{-i k_1 d}}{k^2+q^2} \left(\frac{k_1+ik_2}{\sqrt{k_1^2+k_2^2}}\right)^m J_m(k R_1) \\
&\quad \times\left(\frac{k_1+ik_2}{\sqrt{k_1^2+k_2^2}}\right)^{-m'} J_{m'}(k R_2) \frac{d^2k}{2\pi} \, ,
\end{split}
\end{equation}
where $k=\sqrt{k_1^2+k_2^2}$. The limits of integration are, of
course, $0$ to $2 \pi$ for the angles $\phi$ and $\phi'$ and $-\infty$
to $\infty$ for the $k_1$ and $k_2$ integrals. Despite the appearance
of square roots the above expression is analytical in the integration
variables except for the simple poles due to $k_1^2 + k_2^2 + q^2$ in
the denominator. So, the $k_1$ integration can be carried out by
contour integration. If $d$ is positive then the contour is closed in
the lower half plane. The result is Eq.~\eqref{eq:M23-D}:
\begin{equation}
\begin{split}
& \mbra \tM_{23} \mpket = \\
& (-i)^{m+m'} I_m(q R_1) I_{m'}(q R_2) \int \frac{dk_2}{2 \sqrt{q^2+k_2^2}}
e^{-d\sqrt{q^2+k_2^2}}  \\
& \times\left(\frac{\sqrt{q^2 + k_2^2}+k_2}{q}\right)^{m-m'} = \\
& (-i)^{m+m'} I_m(q R_1) I_{m'}(q R_2) K_{m-m'}(q d).
\end{split}
\end{equation}

$\tM_{23}$ for TE (Neumann) modes is easy to obtain. The derivatives
in Eq.~\eqref{eq:N-def-elements} can be taken out of the integral -they
are just derivatives with respect to $R_1$ and $R_2$- and applied to
$\tM_{23}$ for TM (Dirichlet) modes.

\section{Inverse matrix and matrix elements for multibody forces}
\label{app:multibody}

To compute forces according to Eq.~\eqref{eq:FDN} the matrix $M$ or its
Fourier transform $\tM$ need to be inverted. For a $2\times 2$ block matrix
the inverse can be written in terms of the inverses of the blocks,
\begin{equation}
\begin{split}
&\left(
\begin{array}{cc}
A & B \\
C & D
\end{array}
\right)\inv =\\
&
\left(
\begin{array}{cc}
(A-BD\inv C)\inv & -A\inv B(D-CA\inv B)\inv \\
-D\inv C(A-BD\inv C)\inv & (D-CA\inv B)\inv
\end{array}
\right) \, .
\end{split}
\end{equation}
So, with $\tMnp = [\tM_{\text{np},22},\tM_{\text{np},23};
\tM_{\text{np},32},\tM_{\text{np},33}]$ given, $\tMnp\inv$ can be
found, then the trace taken, to obtain the force.

To express $\tMnp$ in terms of the matrix elements of $\tM_{22}$,
$\tM_{2(2,R)}$, $\tM_{23}$, and $\tM_{2(3,R)}$ we shall refer to their
parameters in the following way
\begin{equation}
\begin{split}
&\tM_{2(2,R)}=\tM_{2(2,R)}(\delta,\mp) \\
&\tM_{23}=\tM_{23}(d) \\
&\tM_{2(3,R)}=\tM_{2(3,R)}(d,\delta,\mp) \, ,
\end{split}
\end{equation}
where the parameters refer to the notation in Eqs.~\eqref{eq:M23-D}, \eqref{eq:M2-2R-D}.
The matrix blocks of $\tMonep$ are built up as follows:
\begin{equation}
\begin{split}
\tM_{\text{1p},22} &= \tM_{22}-s^X \tM_{2(2,R)}(2H,-) \\
\tM_{\text{1p},23} &= \tM_{23}(d)-s^X \tM_{2(3,R)}(d,2H,-).
\end{split}
\end{equation}
Clearly, $\tM_{\text{1p},33}$ is obtained by replacing $R_1$ by $R_2$
and $\tM_{\text{1p},32} = \tM_{\text{1p},23}^\dagger$. As before, one
takes $s^X=+1$ for Dirichlet and $s^X=-1$ for Neumann modes. Because
the reflected cylinders are reflected only once, the minus sign is
chosen in Eq.~\eqref{eq:M2-2R-D}, as indicated by the minus
sign in the arguments above.

$\tMtwop$ can be expressed in as follows
\begin{equation}
\begin{split}
\tM_{\text{2p},22} = & \tM_{22} \\
-s^X \sum_{n=0}^\infty &\left[ \tM_{2(2,R)}(2H_1\!+\!2n(H_1\!+\!H_2),-) \right. \\
+ & \left. \tM_{2(2,R)}(-2H_2\!-\!2n(H_1\!+\!H_2),-)\right]\\
+ \sum_{n=1}^\infty &\left[ \tM_{2(2,R)}(2n(H_1\!+\!H_2),+) \right. \\
+ & \left. \tM_{2(2,R)}(-2n(H_1\!+\!H_2),+)\right]\\
\tM_{\text{2p},23} = & \tM_{23}(d) \\
-s^X \sum_{n=0}^\infty &\left[ \tM_{2(3,R)}(d,2H_1\!+\!2n(H_1\!+\!H_2),-) \right. \\
+ & \left. \tM_{2(3,R)}(d,-2H_2\!-\!2n(H_1\!+\!H_2),-)\right]\\
+ \sum_{n=1}^\infty &\left[ \tM_{2(3,R)}(d,2n(H_1\!+\!H_2),+) \right. \\
+ & \left. \tM_{2(3,R)}(d,-2n(H_1\!+\!H_2),+)\right] \, .
\end{split}
\end{equation}
Again, $\tM_{\text{2p},33}$ is obtained by replacing $R_1$ by $R_2$
and $\tM_{\text{2p},32} = \tM_{\text{2p},23}^\dagger$. Now cylinders
can be reflected an arbitrary number of times, thus, for an odd number
of reflections the minus sign is chosen in Eq.~\eqref{eq:M2-2R-D} and
for an even number of reflections the plus sign.

\section{Asymptotic expansion of the force between two cylinders in the presence of one plate}
\label{app:asymp}
For simplicity, we consider the case of one sidewall and
cylinders of equal radius $R$.  The logarithm of the determinant of
the matrix of Eq.~\eqref{eq:M_np} with $n=1$ can be expressed as
\begin{equation}
\begin{split}
 \label{eq:log-expansion}
&  \log \det(\tilde M_{1p} \tilde M_{1p,\infty}^{-1}) \\
& =  -\sum_{p=1}^\infty \frac{1}{p} \text{Tr} 
(\tilde M_{1p,33}^{-1} \tilde M_{1p,32}  
\tilde M_{1p,22}^{-1} \tilde M_{1p,23})^p \\
& + \text{d-independent terms} \, ,
\end{split}
\end{equation}
where we have used  $\log\det = \text{Tr} \log$ and expanded
the logarithm. Due to the presence of the sidewall, the 
``self-energy'' matrices with elements (see Appendix \ref{app:multibody})
\begin{equation}
\begin{split}
 \label{eq:self-energy-matrices_D}
&  \mbra  \tilde M_{1p,22}\mpket= \mbra  \tilde M_{1p,33}\mpket\\
& = \delta_{mm'} I_m(Rq) K_m(Rq) - I_m(Rq)I_{m'}(Rq) K_{m+m'}(2Hq)
\end{split}
\end{equation}
for D modes and
\begin{equation}
\begin{split}
 \label{eq:self-energy-matrices_N}
&  \mbra \tilde M_{1p,22} \mpket= \mbra  \tilde M_{1p,33}\mpket\\
& = q^2\left[ \delta_{mm'} I'_m(Rq) K'_m(Rq) + I'_m(Rq)I'_{m'}(Rq) 
K_{m+m'}(2Hq)\right]
\end{split}
\end{equation}
for N modes are non-diagonal. For $H\gg R$ the non-diagonal part can
be treated as a small perturbation, and the matrix can be inverted
perturbatively,
\begin{equation}
 \label{eq:inverse_self_energy_matrix}
  M_{1p,22}^{-1} =  M_{1p,22,\infty}^{-1}  \sum_{n=0}^\infty (-1)^n \tilde N^n
\end{equation}
where $M_{1p,22,\infty}^{-1}$ is the diagonal part of
Eqs. \eqref{eq:self-energy-matrices_D},
\eqref{eq:self-energy-matrices_N} and $\tilde N$ is given by
\begin{equation}
 \tilde N = -\frac{I_m(Rq)}{K_{m'}(Rq)}K_{m+m'}(2Hq)
\end{equation}
for D modes and
\begin{equation}
 \tilde N = \frac{I'_m(Rq)}{K'_{m'}(Rq)}K_{m+m'}(2Hq)
\end{equation}
for N modes. We shall see below that the series of
Eq.~\eqref{eq:inverse_self_energy_matrix} yields a rapidly converging
series for the Casimir interaction between the cylinders for all
sidewall separations $H\ge R$. 
Since $d$ is the largest length scale, it is useful to set $q=u/d$ in
Eq.~\eqref{eq:EDN2} since this allows for expansions of Bessel and
Hankel functions that simplify further computations. Using
Eq.~\eqref{eq:log-expansion}, we obtain for the $d$-dependent part of
the energy the series
\begin{equation}
\begin{split}
 \label{eq:energy-of-d}
& E^{D/N} = -\frac{\hbar c L}{4\pi d^2} \int_0^\infty \!\!\! u du \sum_{p=1}^\infty
\frac{1}{p} \text{Tr} \left[
M_{1p,33,\infty}^{-1} 
\left(\sum_{n=0}^\infty (-\tilde N)^n \right) \right. \\
& \left. \times \tilde M_{1p,32} M_{1p,22,\infty}^{-1} \left(\sum_{n=0}^\infty (-\tilde N)^n \right)
\tilde M_{1p,23}
\right]^p \, .
\end{split}
\end{equation}
To leading order in $1/d$, the integral scales as $1/d^4$. At this
order it is sufficient to consider the term for $p=1$ only. Counting
the powers of $1/d$ in the matrix elements appearing in
Eq.~\eqref{eq:energy-of-d} shows that at order $1/d^4$ it is
sufficient to truncate the matrices at order $m=1$ for D modes and
$m=3$ for N modes.  To leading order in $1/d$ the matrix elements of
$\tilde N$ are $\sim (R/H)^{|m|+|m'|}$ and hence independent of $d$ so
that formally all terms of the series over $n$ have to be
included. However, since $R<H$ the matrix elements of $\tilde N^n$
decrease with increasing $n$, yielding a rapidly converging series for
the energy. After the expansion in $1/d$ the integration over $u$ in
Eq.~\eqref{eq:energy-of-d} can be easily performed for N modes since
the integrand consists of terms that are powers over $u$ and the
Hankel functions $K_m(u)$. However, for D modes, there are logarithmic
corrections of the form $\log(Hu/d)$ and $1/\log(Ru/2d)$. For large
$d$, we can make the approximations $\log(Hu/d) \to \log(H/d)$ and
$\log(Ru/2d)\to \log(R/2d)$ since the main contribution to the
integral comes form $u$ of order unity.  With these approximations the
integration over $u$ can be performed analytically.
By including terms up to $n=5$ in the series of Eq.~\eqref{eq:energy-of-d},
we obtain for the force between the cylinders from N modes (with $r=R/H$) 
\begin{equation}
\begin{split}
\label{eq:N-force}
&  \frac{F_N}{L} = -\frac{148}{5\pi} \hbar c \frac{R^4}{d^7} \left[
1+ \frac{29}{74} r^2 + \frac{81}{592} 
r^4 \right. \\
& \left. + \frac{55}{592} 
r^6 + \frac{625}{9472} r^8 
+\frac{201}{4736}r^{10} +
\ldots 
\right] + \ldots,
\end{split}
\end{equation}
to leading order in $1/d$. For D modes, the general form of the
coefficients of the series in $r$ and its first logarithmic
correction can be conjectured from the first few terms. Resummation
yields then the closed form expression for the force for all $r$
to leading order in $1/d$,
\begin{equation}
\begin{split}
  \label{eq:D-force}
  \frac{F_D}{L} & = -\frac{\hbar c}{\pi} \frac{R^4}{d^7} \bigg[ \frac{64}{(4-r^2)^2}
-\frac{64}{\log(R/2d)} \frac{(4-3r^2)^2}{r^2(4-r^2)^3} \\
& + F_2(r) \log^{-2}(R/2d) +  \, {\cal O}(\log^{-3}(R/2d)) \bigg] + \ldots \, .
\end{split}
\end{equation}
Here $F_2(r) = 4/r^4$ for $r\to 0$ so that for $H\gg R$ the force
is proportional to $H^4/(d^7 \log^2(R/2d))$, which is the same as
Eq. (\ref{asymptote}) up to the different $\log$ corrections. Notice that the
expansion in $1/\log(R/2d)$ of Eq.~\eqref{eq:D-force} is formal and
convergence for all $r$ is not assured in the Dirichlet case. The
reason for that is related to logarithmic corrections $\sim
\log(H/d)=\log(R/2d)+\log(2/r)$ to the matrix elements of $\tilde N$
so that the expansion coefficients grow with $H$ in a way that higher
order coefficients involve higher powers of $\log(H/d)$.

The total force is given by the sum $F_N+F_D$. From this large
distance expansion we can obtain insight into the generation of the
non-monotonic behavior of the force.  The Neumann force expansion is
simple to understand, $F_N$ decreases in magnitude as the cylinders
move away from the plate, i.e., $r = R/H$ decreases from $1$.
Surprisingly, the expansion of the Dirichlet force {\it without} the
logarithmic terms of Eq.~\eqref{eq:D-force}, i.e. at very large
cylinder-cylinder separations $d$, indicates that $F_D$, similar to
$F_N$, decreases as the cylinders move away from the plate. But
ultimately, as $H$ increases, the coefficients of the inverse $\log$
terms dominate and $F_D$ increases as the cylinders move away, albeit
at a slower rate compared to $F_N$, as expected from our numerical
results and previous considerations in terms of image charges. If
valid, therefore, the expansion not only captures the opposing changes
in $F_D$ and $F_N$ with sidewall separation but also suggests that
$F_D$ \emph{itself} has interesting non-monotonic behavior.

\end{appendix}

\bibliographystyle{apsrev}
\bibliography{../article}

\begin{thebibliography}{22}
\expandafter\ifx\csname natexlab\endcsname\relax\def\natexlab#1{#1}\fi
\expandafter\ifx\csname bibnamefont\endcsname\relax
  \def\bibnamefont#1{#1}\fi
\expandafter\ifx\csname bibfnamefont\endcsname\relax
  \def\bibfnamefont#1{#1}\fi
\expandafter\ifx\csname citenamefont\endcsname\relax
  \def\citenamefont#1{#1}\fi
\expandafter\ifx\csname url\endcsname\relax
  \def\url#1{\texttt{#1}}\fi
\expandafter\ifx\csname urlprefix\endcsname\relax\def\urlprefix{URL }\fi
\providecommand{\bibinfo}[2]{#2}
\providecommand{\eprint}[2][]{\url{#2}}

\bibitem[{\citenamefont{Emig et~al.}(2006)\citenamefont{Emig, Jaffe, Kardar,
  and Scardicchio}}]{Emig06}
\bibinfo{author}{\bibfnamefont{T.}~\bibnamefont{Emig}},
  \bibinfo{author}{\bibfnamefont{R.~L.} \bibnamefont{Jaffe}},
  \bibinfo{author}{\bibfnamefont{M.}~\bibnamefont{Kardar}}, \bibnamefont{and}
  \bibinfo{author}{\bibfnamefont{A.}~\bibnamefont{Scardicchio}},
  \bibinfo{journal}{Physical Review Letters} \textbf{\bibinfo{volume}{96}},
  \bibinfo{pages}{080403} (\bibinfo{year}{2006}).

\bibitem[{\citenamefont{Cleland and Roukes}(1996)}]{Cleland96}
\bibinfo{author}{\bibfnamefont{A.~N.} \bibnamefont{Cleland}} \bibnamefont{and}
  \bibinfo{author}{\bibfnamefont{M.~L.} \bibnamefont{Roukes}},
  \bibinfo{journal}{Appl. Phys. Lett.} \textbf{\bibinfo{volume}{69}},
  \bibinfo{pages}{2653} (\bibinfo{year}{1996}).

\bibitem[{\citenamefont{Chan et~al.}(2001)\citenamefont{Chan, Aksyuk, Kleiman,
  Bishop, and Capasso}}]{Chan01}
\bibinfo{author}{\bibfnamefont{H.~B.} \bibnamefont{Chan}},
  \bibinfo{author}{\bibfnamefont{V.~A.} \bibnamefont{Aksyuk}},
  \bibinfo{author}{\bibfnamefont{R.~N.} \bibnamefont{Kleiman}},
  \bibinfo{author}{\bibfnamefont{D.~J.} \bibnamefont{Bishop}},
  \bibnamefont{and} \bibinfo{author}{\bibfnamefont{F.}~\bibnamefont{Capasso}},
  \bibinfo{journal}{Science} \textbf{\bibinfo{volume}{291}},
  \bibinfo{pages}{1941} (\bibinfo{year}{2001}).

\bibitem[{\citenamefont{Sazonova et~al.}(2004)\citenamefont{Sazonova, Yaish,
  {\"U}st{\"u}nel, Roundy, Arias, and McEuen}}]{Sazonova04}
\bibinfo{author}{\bibfnamefont{V.}~\bibnamefont{Sazonova}},
  \bibinfo{author}{\bibfnamefont{Y.}~\bibnamefont{Yaish}},
  \bibinfo{author}{\bibfnamefont{H.}~\bibnamefont{{\"U}st{\"u}nel}},
  \bibinfo{author}{\bibfnamefont{D.}~\bibnamefont{Roundy}},
  \bibinfo{author}{\bibfnamefont{T.~A.} \bibnamefont{Arias}}, \bibnamefont{and}
  \bibinfo{author}{\bibfnamefont{P.~L.} \bibnamefont{McEuen}},
  \bibinfo{journal}{Nature} \textbf{\bibinfo{volume}{431}},
  \bibinfo{pages}{284} (\bibinfo{year}{2004}).

\bibitem[{\citenamefont{Casimir}(1948)}]{Casimir48-1}
\bibinfo{author}{\bibfnamefont{H.~B.~G.} \bibnamefont{Casimir}},
  \bibinfo{journal}{Proc. K. Ned. Akad. Wet.} \textbf{\bibinfo{volume}{51}},
  \bibinfo{pages}{793} (\bibinfo{year}{1948}).

\bibitem[{\citenamefont{Hertzberg et~al.}(2005)\citenamefont{Hertzberg, Jaffe,
  Kardar, and Scardicchio}}]{Hertzberg05}
\bibinfo{author}{\bibfnamefont{M.~P.} \bibnamefont{Hertzberg}},
  \bibinfo{author}{\bibfnamefont{R.~L.} \bibnamefont{Jaffe}},
  \bibinfo{author}{\bibfnamefont{M.}~\bibnamefont{Kardar}}, \bibnamefont{and}
  \bibinfo{author}{\bibfnamefont{A.}~\bibnamefont{Scardicchio}},
  \bibinfo{journal}{Phys. Rev. Lett.} \textbf{\bibinfo{volume}{95}},
  \bibinfo{pages}{250402} (\bibinfo{year}{2005}).

\bibitem[{\citenamefont{Casimir and Polder}(1948)}]{Casimir48-2}
\bibinfo{author}{\bibfnamefont{H.~B.~G.} \bibnamefont{Casimir}}
  \bibnamefont{and} \bibinfo{author}{\bibfnamefont{D.}~\bibnamefont{Polder}},
  \bibinfo{journal}{Phys. Rev.} \textbf{\bibinfo{volume}{73}},
  \bibinfo{pages}{360} (\bibinfo{year}{1948}).

\bibitem[{\citenamefont{Emig}(2007)}]{Emig:2007a}
\bibinfo{author}{\bibfnamefont{T.}~\bibnamefont{Emig}},
  \emph{\bibinfo{title}{Fluctuation induced quantum interactions between
  compact objects and a plane mirror}}, \bibinfo{howpublished}{Preprint
  arXiv:0712.2199} (\bibinfo{year}{2007}).

\bibitem[{\citenamefont{Scardicchio}(2005)}]{Scardicchio:2005b}
\bibinfo{author}{\bibfnamefont{A.}~\bibnamefont{Scardicchio}},
  \bibinfo{journal}{Phys. Rev. D} \textbf{\bibinfo{volume}{72}},
  \bibinfo{pages}{065004} (\bibinfo{year}{2005}),
  \urlprefix\url{http://link.aps.org/abstract/PRD/v72/p065004}.

\bibitem[{\citenamefont{Rodriguez et~al.}(2007)\citenamefont{Rodriguez,
  Ibanescu, Iannuzzi, Capasso, Joannopoulos, and Johnson}}]{Rodriguez07:PRL}
\bibinfo{author}{\bibfnamefont{A.}~\bibnamefont{Rodriguez}},
  \bibinfo{author}{\bibfnamefont{M.}~\bibnamefont{Ibanescu}},
  \bibinfo{author}{\bibfnamefont{D.}~\bibnamefont{Iannuzzi}},
  \bibinfo{author}{\bibfnamefont{F.}~\bibnamefont{Capasso}},
  \bibinfo{author}{\bibfnamefont{J.~D.} \bibnamefont{Joannopoulos}},
  \bibnamefont{and} \bibinfo{author}{\bibfnamefont{S.~G.}
  \bibnamefont{Johnson}}, \bibinfo{journal}{Phys. Rev. Lett.}
  \textbf{\bibinfo{volume}{99}}, \bibinfo{pages}{080401}
  (\bibinfo{year}{2007}).

\bibitem[{\citenamefont{Rahi et~al.}(2008)\citenamefont{Rahi, Rodriguez, Emig,
  Jaffe, Johnson, and Kardar}}]{Rahi:2008mz}
\bibinfo{author}{\bibfnamefont{S.~J.} \bibnamefont{Rahi}},
  \bibinfo{author}{\bibfnamefont{A.~W.} \bibnamefont{Rodriguez}},
  \bibinfo{author}{\bibfnamefont{T.}~\bibnamefont{Emig}},
  \bibinfo{author}{\bibfnamefont{R.~L.} \bibnamefont{Jaffe}},
  \bibinfo{author}{\bibfnamefont{S.~G.} \bibnamefont{Johnson}},
  \bibnamefont{and} \bibinfo{author}{\bibfnamefont{M.}~\bibnamefont{Kardar}},
  \bibinfo{journal}{Phys. Rev. A} \textbf{\bibinfo{volume}{77}},
  \bibinfo{eid}{030101} (pages~\bibinfo{numpages}{4}) (\bibinfo{year}{2008}),
  \urlprefix\url{http://link.aps.org/abstract/PRA/v77/e030101}.

\bibitem[{\citenamefont{Brown-Hayes et~al.}(2005)\citenamefont{Brown-Hayes,
  Dalvit, Mazzitelli, Kim, and Onofrio}}]{Brown-Hayes05}
\bibinfo{author}{\bibfnamefont{M.}~\bibnamefont{Brown-Hayes}},
  \bibinfo{author}{\bibfnamefont{D.~A.~R.} \bibnamefont{Dalvit}},
  \bibinfo{author}{\bibfnamefont{F.~D.} \bibnamefont{Mazzitelli}},
  \bibinfo{author}{\bibfnamefont{W.~J.} \bibnamefont{Kim}}, \bibnamefont{and}
  \bibinfo{author}{\bibfnamefont{R.}~\bibnamefont{Onofrio}},
  \bibinfo{journal}{Phys. Rev. A} \textbf{\bibinfo{volume}{72}},
  \bibinfo{pages}{052102} (\bibinfo{year}{2005}).

\bibitem[{\citenamefont{Li and Kardar}(1991)}]{Li91}
\bibinfo{author}{\bibfnamefont{H.}~\bibnamefont{Li}} \bibnamefont{and}
  \bibinfo{author}{\bibfnamefont{M.}~\bibnamefont{Kardar}},
  \bibinfo{journal}{Phys. Rev. Lett.} \textbf{\bibinfo{volume}{67}},
  \bibinfo{pages}{3275} (\bibinfo{year}{1991}).

\bibitem[{\citenamefont{Li and Kardar}(1992)}]{Li92}
\bibinfo{author}{\bibfnamefont{H.}~\bibnamefont{Li}} \bibnamefont{and}
  \bibinfo{author}{\bibfnamefont{M.}~\bibnamefont{Kardar}},
  \bibinfo{journal}{Phys. Rev. A} \textbf{\bibinfo{volume}{46}},
  \bibinfo{pages}{6490} (\bibinfo{year}{1992}).

\bibitem[{\citenamefont{Buescher and Emig}(2005)}]{Buescher05}
\bibinfo{author}{\bibfnamefont{R.}~\bibnamefont{Buescher}} \bibnamefont{and}
  \bibinfo{author}{\bibfnamefont{T.}~\bibnamefont{Emig}},
  \bibinfo{journal}{Physical Review Letters} \textbf{\bibinfo{volume}{94}},
  \bibinfo{pages}{133901} (\bibinfo{year}{2005}).

\bibitem[{\citenamefont{Dalvit et~al.}(2004)\citenamefont{Dalvit, Lombardo,
  Mazzitelli, and Onofrio}}]{Dalvit04}
\bibinfo{author}{\bibfnamefont{D.~A.~R.} \bibnamefont{Dalvit}},
  \bibinfo{author}{\bibfnamefont{F.~C.} \bibnamefont{Lombardo}},
  \bibinfo{author}{\bibfnamefont{F.~D.} \bibnamefont{Mazzitelli}},
  \bibnamefont{and} \bibinfo{author}{\bibfnamefont{R.}~\bibnamefont{Onofrio}},
  \bibinfo{journal}{Europhys. Lett.} \textbf{\bibinfo{volume}{67}},
  \bibinfo{pages}{517} (\bibinfo{year}{2004}).

\bibitem[{\citenamefont{Kenneth et~al.}(2002)\citenamefont{Kenneth, Klich,
  Mann, and Revzen}}]{Kenneth02}
\bibinfo{author}{\bibfnamefont{O.}~\bibnamefont{Kenneth}},
  \bibinfo{author}{\bibfnamefont{I.}~\bibnamefont{Klich}},
  \bibinfo{author}{\bibfnamefont{A.}~\bibnamefont{Mann}}, \bibnamefont{and}
  \bibinfo{author}{\bibfnamefont{M.}~\bibnamefont{Revzen}},
  \bibinfo{journal}{Phys. Rev. Lett.} \textbf{\bibinfo{volume}{89}},
  \bibinfo{pages}{033001} (\bibinfo{year}{2002}).

\bibitem[{\citenamefont{Imry}(2005)}]{Imry05}
\bibinfo{author}{\bibfnamefont{Y.}~\bibnamefont{Imry}}, \bibinfo{journal}{Phys.
  Rev. Lett.} \textbf{\bibinfo{volume}{95}}, \bibinfo{pages}{080404}
  (\bibinfo{year}{2005}).

\bibitem[{\citenamefont{Munday and Capasso}(2007)}]{Munday07}
\bibinfo{author}{\bibfnamefont{J.~N.} \bibnamefont{Munday}} \bibnamefont{and}
  \bibinfo{author}{\bibfnamefont{F.}~\bibnamefont{Capasso}},
  \bibinfo{journal}{Phys. Rev. A} \textbf{\bibinfo{volume}{75}},
  \bibinfo{pages}{060102(R)} (\bibinfo{year}{2007}).

\bibitem[{\citenamefont{Leonhardt and Philbin}(2007)}]{Leonhardt07}
\bibinfo{author}{\bibfnamefont{U.}~\bibnamefont{Leonhardt}} \bibnamefont{and}
  \bibinfo{author}{\bibfnamefont{T.~G.} \bibnamefont{Philbin}},
  \bibinfo{journal}{New J. Phys.} \textbf{\bibinfo{volume}{9}},
  \bibinfo{pages}{254} (\bibinfo{year}{2007}).

\bibitem[{\citenamefont{Emig et~al.}(2007)\citenamefont{Emig, Graham, Jaffe,
  and Kardar}}]{Emig07}
\bibinfo{author}{\bibfnamefont{T.}~\bibnamefont{Emig}},
  \bibinfo{author}{\bibfnamefont{N.}~\bibnamefont{Graham}},
  \bibinfo{author}{\bibfnamefont{R.~L.} \bibnamefont{Jaffe}}, \bibnamefont{and}
  \bibinfo{author}{\bibfnamefont{M.}~\bibnamefont{Kardar}},
  \bibinfo{journal}{Phys. Rev. Lett.} \textbf{\bibinfo{volume}{99}},
  \bibinfo{pages}{170403} (\bibinfo{year}{2007}).

\bibitem[{\citenamefont{Emig et~al.}(2008)\citenamefont{Emig, Graham, Jaffe,
  and Kardar}}]{Emig08}
\bibinfo{author}{\bibfnamefont{T.}~\bibnamefont{Emig}},
  \bibinfo{author}{\bibfnamefont{N.}~\bibnamefont{Graham}},
  \bibinfo{author}{\bibfnamefont{R.~L.} \bibnamefont{Jaffe}}, \bibnamefont{and}
  \bibinfo{author}{\bibfnamefont{M.}~\bibnamefont{Kardar}},
  \bibinfo{journal}{Phys. Rev. D} \textbf{\bibinfo{volume}{77}},
  \bibinfo{pages}{025005} (\bibinfo{year}{2008}).

\end{thebibliography}

\end{document}